\documentclass[aps,prl,twocolumn,superscriptaddress]{revtex4}
\pdfoutput=1

\usepackage{graphicx}
\usepackage{amsmath}
\usepackage{amssymb}
\usepackage{color}
\usepackage{amscd}
\usepackage{bm}

\DeclareMathOperator{\atan}{atan}

\begin{document}
\title{Statistical Mechanics of Developable Ribbons}

\author{L. Giomi}
\affiliation{School of Engineering and Applied Sciences, Harvard University, Cambridge, MA 02138, USA}
\affiliation{Martin A. Fisher School of Physics, Brandeis University, Waltham, MA 02454, USA}

\author{L. Mahadevan}
\affiliation{School of Engineering and Applied Sciences, Harvard University, Cambridge, MA 02138, USA}

\date{\today}

\begin{abstract}
\noindent We investigate the statistical mechanics of long developable ribbons of finite width and very small thickness. The constraint of isometric 
deformations in these ribbon-like structures that follows from the geometric separation of scales introduces a coupling between bending and torsional 
degrees of freedom. Using analytical techniques and Monte Carlo simulations, we find that the tangent-tangent correlation functions always exhibits an 
oscillatory decay at any finite temperature implying the existence of an underlying helical structure even in absence of a preferential zero-temperature 
twist. In addition the persistence length is found to be over three times larger than that of a wormlike chain having the same bending rigidity. Our results 
are applicable to many ribbon-like objects in polymer physics and nanoscience that cannot be described by the classical worm-like chain model.

\end{abstract}

\maketitle

\noindent A shell or plate is a body whose two dimensions are comparable to each other and much larger than the third.  A rod, on the other hand, is a body whose two
dimensions are comparable to each other and much smaller than the third. In each of these objects, this separation of scales leads to constraints and an 
energetic ordering of the different possible deformations such as bending,  twisting, stretching and shearing, and there are well accepted continuum and statistical 
theories of their behavior.  There is, finally, an intermediate class of objects where all three dimensions are widely separated from each other, which we call strips or ribbons. 

\begin{figure}[b]
\centering
\includegraphics[width=0.75\columnwidth]{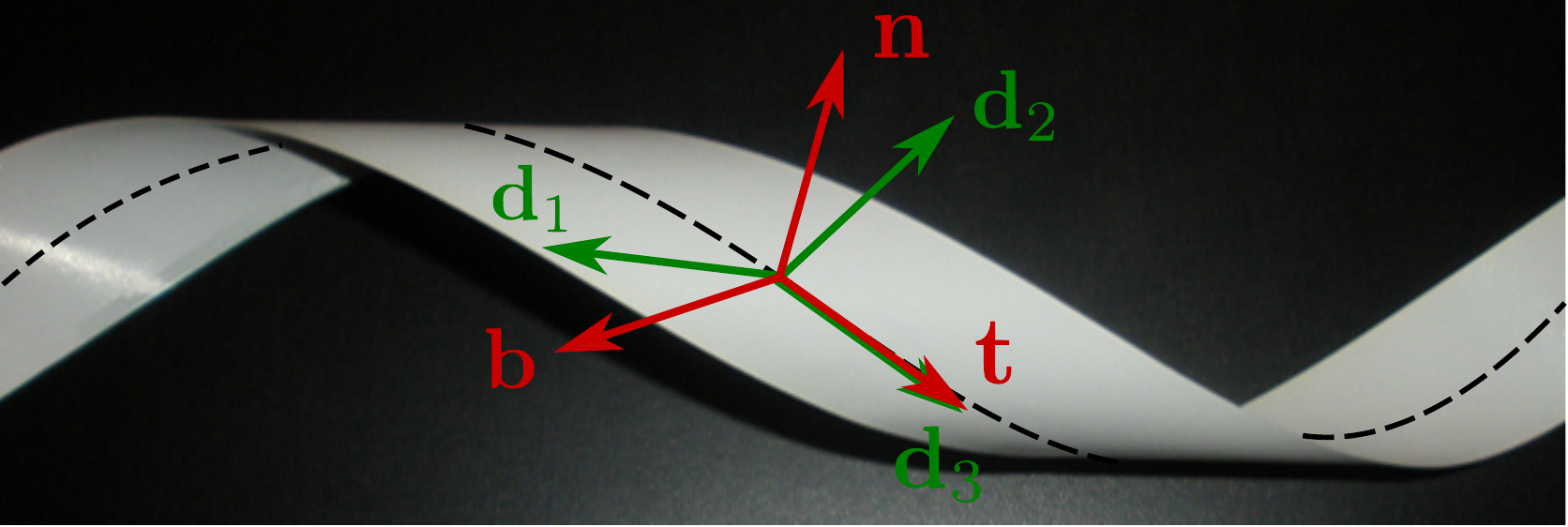}
\caption{\label{fig:photo-ribbon}(Color online) A twisted paper ribbon. In presence of twist the ribbon necessarily develop a non-zero curvature, as required by 
the Sadowsky functional. The Frenet frame $({\bf t},\,{\bf n},\,{\bf b})$ and the material frame $({\bf d}_{1},\,{\bf d}_{2},\,{\bf d}_{3})$ are sketched in red and 
green respectively. For a developable ribbon, the two frames coincide, so that the twist of the material frame and the torsion of the center line are identical.}
\end{figure}

An example of such a chimeric object is a rectangular sheet of paper of large aspect ratio (see Fig. \ref{fig:photo-ribbon}); its thickness is much smaller than its width which 
is itself much smaller than its length. The nanoscale world offers a number of compelling examples of these ribbon-like structures such as DNA, the secondary structures of 
proteins such as $
\alpha$-helices and $\beta$-sheets, Carbon (graphene) and Molybdenum ribbons etc. A classical theoretical approach to the study of these filamentous objects is embodied 
in the wormlike chain (WLC) model \cite{KratkyPorod:1949}, in which the polymer is assumed to have an inextensible backbone, but with a flexibility governed by the 
persistence length $\ell_{p}$ which expresses the exponential rate of decay of tangent-tangent correlation function. Its relative simplicity and precision in describing the 
entropic elasticity of many semiflexible polymers has made it very attractive as an object of theoretical study. However,  the accuracy of the WLC model for finite width 
biopolymers is still debated and a number of alternatives have been proposed  to account for the torsional rigidity, the excluded volume and other features associated with a 
finite width \cite{Polymers,Liverpool,Marenduzzo:2005}. In this Letter we consider a model for the statistical mechanics of ribbon-like objects based on a classical geometric 
formulation of developable surfaces. In the limit when the thickness of a ribbon is much less than its width which is itself much less than its length, we can define a 
centerline. Then isometric deformations of the ribbon (i.e. those that preserve local distances so that the metric tensor is always Euclidean, $g_{ij}=\delta_{ij}$) can be 
described in terms of the orientation of the generators of the developable surface relative to the centerline \cite{Sadowsky:1930} and account for the finite width of the ribbon. 
Because of its tensorial nature, developability strongly constrains the allowable deformations of a ribbon in  a fundamentally different way than other milder constraints such
as the ``railway-track'' models of polymers \cite{Liverpool}. The most striking consequence of this, illustrated in Fig. \ref{fig:photo-ribbon}, is that a developable ribbon can bend 
without twisting but {\em cannot twist without bending}, which as we shall see has far reaching consequences.

To characterize a ribbon, we first consider the geometry of its centerline in terms of the Frenet frame $({\bf t},\,{\bf n},\,{\bf b}) (s)$ of the tangent, normal and 
binormal vector at each point along the curve parametrized by arc-length coordinate $s$ \cite{Struik:1988}.This orthogonal triad evolves in space according to 
the Frenet-Serret equations:
\[
{\bf t}' = \kappa {\bf n}\,, \qquad
{\bf n}' =-\kappa {\bf t} +\tau {\bf b}\,,\qquad
{\bf b}' =-\tau {\bf n}\,.
\]
where $\kappa, \tau$ are the curvature and torsion of the space curve and the prime denotes the derivatives with respect to $s$. In physical ribbons, there is another 
distinguished frame, the material frame which is attached to each cross-section; the spatial rate of rotation of this material frame is known as the twist $\tau_e$ and is not 
the same as the torsion of the geometric centerline. For example, an arbitrary twisted ribbon can have a straight center line, while a helically wound telephone cord may only 
have torsion and no twist. However, when the ribbon is deformed isometrically, the direction perpendicular to the centerline, which points along the binormal, also points along 
a material axis that corresponds locally to the only direction in which bending is possible, so that Frenet frame and the material frame coincide and $\tau = \tau_e$. 
Equivalently, the projection of the curvature of the centerline on the material frame always vanishes in the infinitely stiff direction.

The elastic energy due to isometric deformations of a naturally flat sheet of thickness $h$ is given by the functional: $F = \frac{1}{2}D\int dA\,(\kappa_{1}+
\kappa_{2})^{2}$, where $\kappa_{1}$ and $\kappa_{2}$ are the principal curvatures, $D=h^{3}E/[12(1-\nu^{2})]$ is the bending rigidity, $E$ the Young 
modulus of the material and $\nu$ its Poisson ratio.  The assumption of developability allows one to reduce the two dimensional integral to a one-
dimensional integral along the centerline of the strip \cite{Wunderlich:1962,Mahadevan}:
\[
F = \frac{1}{2}D\int_{0}^{L} ds\,\frac{(\kappa^{2}+\tau^{2})^{2}}{\kappa'\tau-\kappa\tau'}\log\left[\frac{\kappa^{2}+\frac{w}{2}(\kappa'\tau-\kappa\tau')}
{\kappa^{2}-\frac{w}{2}(\kappa'\tau-\kappa\tau')}\right]\,,
\]
where $\kappa$ and $\tau$ are respectively the curvature and the torsion of the centerline, $w$ is the lateral width of the ribbon and $L$ the length of the 
centerline. In the limit $w/L\ll 1$ this yields the Sadowsky functional
\cite{Sadowsky:1930}:
\begin{equation}\label{eq:sadowsky}
F = \frac{1}{2}Dw\int_{0}^{L}ds\,\frac{(\kappa^{2}+\tau^{2})^{2}}{\kappa^{2}}\,. 	
\end{equation}
As anticipated, the nonlinear coupling between curvature and torsion in \eqref{eq:sadowsky} implies that an inextensible ribbon can bend without twisting ($\tau = 0$, $\kappa \ne 0$), but cannot twist without bending ($\tau \ne 0 \rightarrow \kappa \ne 0$).

\begin{figure}[t]
\centering
\includegraphics[scale=0.5]{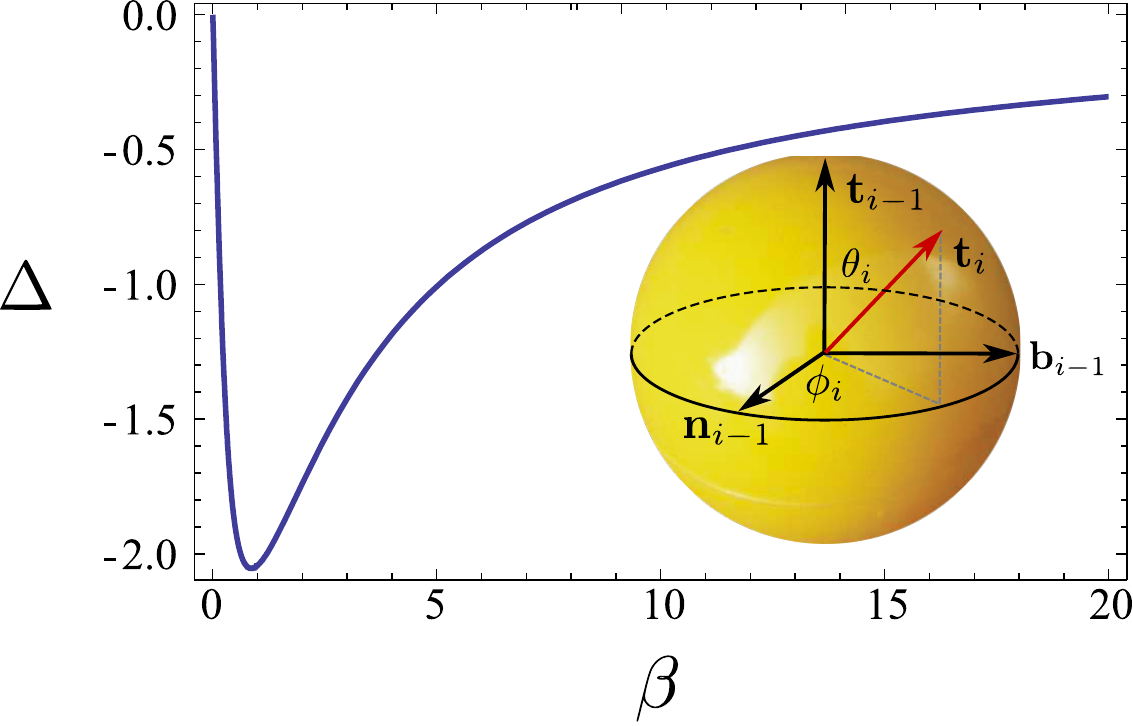}	
\caption{\label{fig:delta}(Color online) The quantity $\Delta$ defined in the text as a function of $\beta$, calculated from the exact integral expression given on the last page. With exception for the two limits $\beta\rightarrow 0,\,\infty$, $\Delta$ is always negative and the associated eigenvalues $\lambda_{1,2}=\frac{1}{2}(b\pm\sqrt{\Delta})$ are complex and conjugate. In the inset the discrete Frenet frame.}
\end{figure}

\begin{figure}[t!]
\centering
\includegraphics[width=.95\columnwidth]{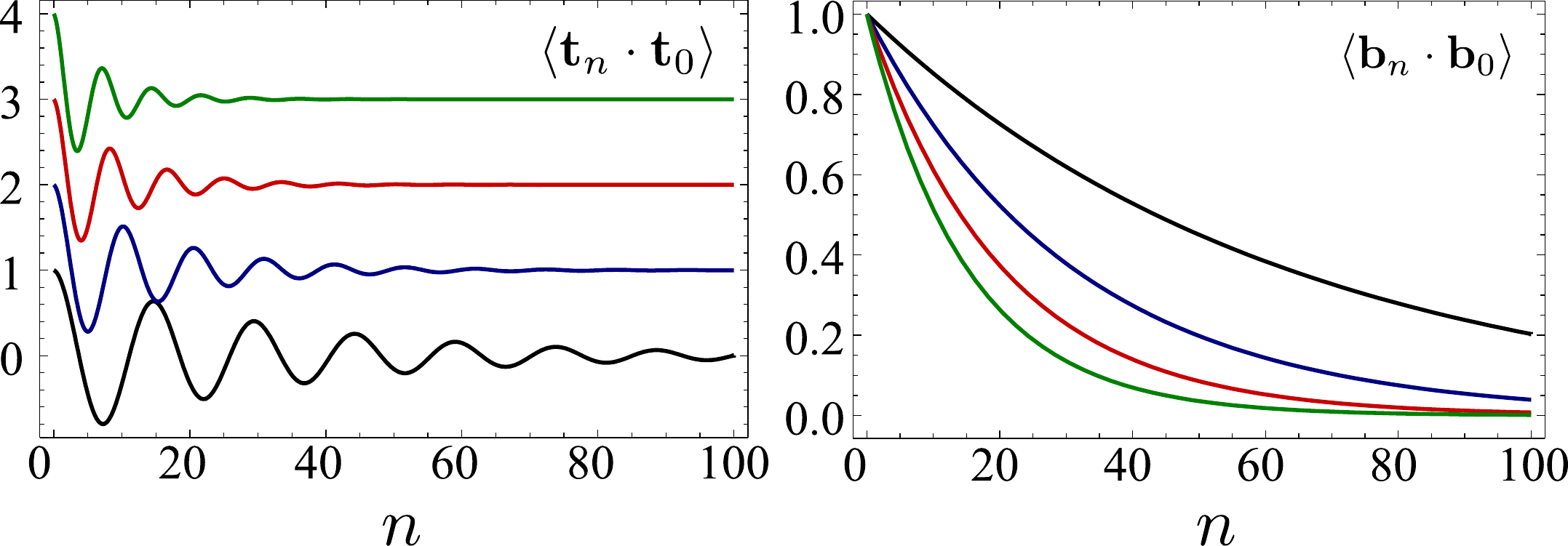}
\caption{\label{fig:correlation-function}(Color online) Tangent-tangent (left) and binormal-binormal (right) correlation functions from Monte Carlo simulations of a 
polymer chain of $N=100$ segments at $T=0.1$ (black), $0.2$ (blue), $0.3$ (red) and $0.4$ (green). The graphs of $\langle {\bf t}_{n}\cdot{\bf 
t}_{0}\rangle$ have been translated along the vertical axis to avoid overlaps.}
\end{figure}

%
%
To make the Sadowsky's functional \eqref{eq:sadowsky} amenable to numerical simulation of a discretized analog, we  consider a chain of $N=L/a$ consecutive vertices $
{\bf x}=\{{\bf x}_{1},\,{\bf x}_{2}\ldots\,{\bf x}_{N}\}$ separated by a fixed length $a$. At each point ${\bf x}_{i}$ in the chain, the discrete Frenet frame is given 
by: ${\bf F}_{i}=({\bf t}_{i},\,{\bf n}_{i},\,{\bf b}_{i})$.
%
%
${\bf F}_{i}$ are orthogonal $3\times 3$ matrices whose columns are given by the column vectors ${\bf t}_{i}$, ${\bf n}_{i}$ and ${\bf b}_{i}$. Calling $\theta_{i}$ and $\phi_{i}$ 
the angles formed by consecutive tangent and binormal vectors (i.e. ${\bf t}_{i} \cdot{\bf t}_{i-1}=\cos\theta_{i}$ and ${\bf b}_{i}\cdot{\bf b}_{i-1}=\cos\phi_{i}$), the Frenet 
equations for the moving trihedron are expressed as ${\bf F}_{i}={\bf F}_{i-1}{\bf R}_{i}$ \cite{Quine:2004}, where
\begin{equation}\label{eq:matrix}
{\bf R}_{i} = 
\left(
\begin{array}{ccc}
\cos\theta_{i} & -\sin\theta_{i}\phantom{-} & 0 \\
\sin\theta_{i}\cos\phi_{i} & \cos\theta_{i}\cos\phi_{i} & -\sin\phi_{i} \\
\sin\theta_{i}\sin\phi_{i} & \cos\theta_{i}\sin\phi_{i} &  \phantom{-}\cos\phi_{i} 
\end{array}
\right)\,.
\end{equation}
We note that ${\bf t}_{i}\cdot {\bf n}_{i-1}=\sin\theta_{i}\cos\phi_{i}$ and ${\bf t}_{i}\cdot {\bf b}_{i-1}=\sin\theta_{i}\sin\phi_{i}$. Thus $\theta_{i}$ and $\phi_{i}
$ are also the polar and azimuthal angle of a system of spherical coordinates (see inset of Fig. \ref{fig:delta}). The discrete analog of curvature and torsion 
can be easily found $\kappa_{i}^{2} = a^{-2}|{\bf t_{i}}-{\bf t}_{i-1}|^{2} = 2a^{-2}(1-\cos\theta_{i})$ and $\tau_{i}^{2} = a^{-2}|{\bf b_{i}}-{\bf b}_{i-1}|^{2} = 2a^{-2}
(1-\cos\phi_{i})$. The energy \eqref{eq:sadowsky} can thus be written as:
\begin{equation}\label{eq:discrete-sadowsky}
\frac{F}{k_{B}T} = \beta \sum_{i=1}^{N} \frac{[(1-{\bf t}_{i}\cdot{\bf t}_{i+1})+(1-{\bf b}_{i}\cdot{\bf b}_{i+1})]^{2}}{(1-{\bf t}_{i}\cdot{\bf t}_{i+1})}\,,
\end{equation}
where $\beta=Dw/(ak_{B}T)$.  This serves as a starting point for Monte Carlo simulations of a  chain composed of $N$ segments of unit length and its conformation is updated using pivot moves through a Metropolis algorithm. For every chain we perform $10^{6}$ Monte Carlo sweeps, the first half of which is devoted to equilibration. Fig. \ref{fig:correlation-function} shows the tangent-tangent and binormal-binormal correlation for a chain of $N=100$ segments at various temperatures. We see that while $
\langle {\bf b}_{n}\cdot{\bf b}_{0} \rangle$ always exhibits exponential decay, $\langle {\bf t}_{n}\cdot{\bf t}_{0} \rangle$ has oscillations superimposed on an 
exponential decay. Upon raising the temperature, the frequency increases, but the exponential damping becomes stronger.

To understand this result, we first derive the tangent-tangent and binormal-binormal correlation functions from the iterated transfer matrix: $\langle {\bf t}_{n}
\cdot{\bf t}_{0}\rangle = \langle {\bf R}_{1}{\bf R}_{2}\ldots\,{\bf R}_{n} \rangle_{11}$ and $\langle {\bf b}_{n}\cdot{\bf b}_{0}\rangle = \langle {\bf R}_{1}{\bf 
R}_{2}\ldots\,{\bf R}_{n} \rangle_{33}$ \cite{Marenduzzo:2005}, noting that since the energy \eqref{eq:discrete-sadowsky} is a sum over individual segments, the Boltzmann 
measure $\exp(-F/k_{B}T)$ is factorizable and $\langle {\bf R}_{1}{\bf R}_{2}\ldots\,{\bf R}_{n} \rangle = \langle {\bf R} \rangle^{n}$. Moreover, for sufficiently large spatial 
separation the behavior of the correlation functions will be dictated by the largest eigenvalue in the spectrum of ${\bf R}$. Assuming $\langle \sin\phi \rangle = 0$, the transfer 
matrix ${\bf R}$ becomes block-diagonal and the eigenvalues are given by the roots of the characteristic equation $\lambda^{2}-b\lambda+c=0$ with $b= \langle 
\cos\theta \rangle + \langle \cos\theta \cos\phi \rangle$ and $c = \langle \cos\theta \rangle \langle \cos\theta \cos\phi \rangle + \langle \sin\theta \rangle \langle \sin \theta \cos
\phi\rangle$, so that
\[
\lambda_{1,2} = \frac{b\pm\sqrt{\Delta}}{2}\,, \qquad
\lambda_{3} = \langle \cos\phi \rangle\,,
\]
where $\Delta=b^{2}-4c$. Thus the binormal-binormal correlation function always exhibits exponential decay: 
\begin{equation}
\langle {\bf b}_{n}\cdot{\bf b}_{0}\rangle = e^{-s/\ell_{\tau}}\,,
\end{equation}
where $\ell_{\tau}^{-1}=-a^{-1}\log\langle \cos\phi \rangle$ is the {\em torsional persistence length} and $s=na$. On the other hand, the sign of $\Delta$ 
dictates the behavior of the tangent-tangent correlation function. When $\Delta>0$, the eigenvalues $\lambda_{1,2}$ are real and $\langle {\bf t}_{n}
\cdot{\bf t}_{0} \rangle$ has the typical exponential decay of a WLC. When $\Delta<0$ the eigenvalues are complex and $\langle {\bf t}_{n}\cdot{\bf t}_{0} 
\rangle$ has the form:
\begin{equation}
\langle {\bf t}_{n}\cdot{\bf t}_{0} \rangle = e^{-s/\ell_{p}}\cos ks\,,
\end{equation}
with $k=a^{-1}\atan(\sqrt{-\Delta}/b)$ and $\ell_{p}^{-1}=-\frac{1}{2}a^{-1}\log c$. An exact numerical calculation of $\Delta$, shown in Fig. \ref{fig:delta} as function of 
$\beta$, shows it is always negative so that the tangent-tangent correlation function is always oscillatory at any non-zero temperature.  

\begin{figure}[t]
\centering
\includegraphics[width=0.8\columnwidth]{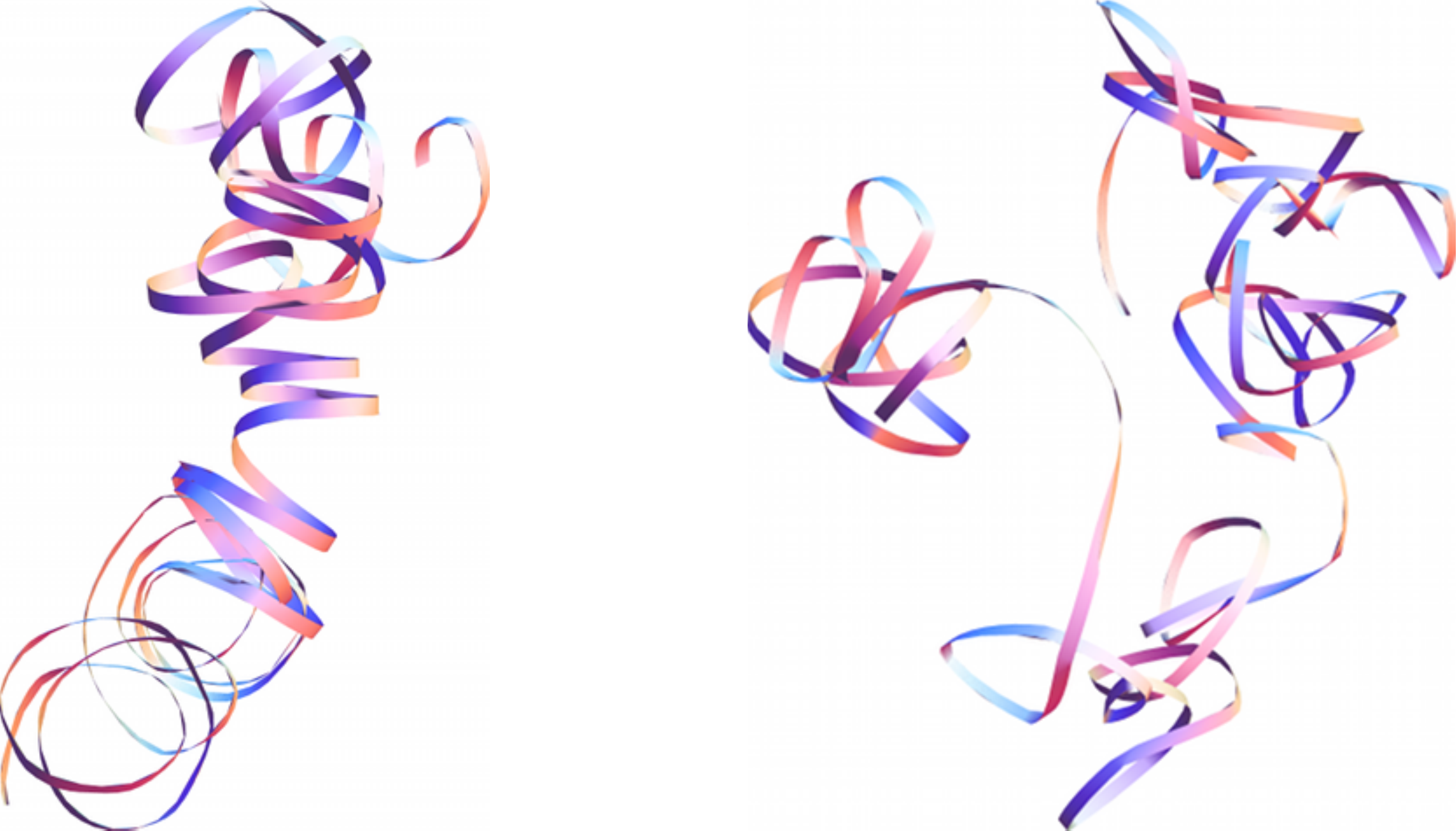}	
\caption{\label{fig:ribbon}(Color online) The typical conformation of a ribbon-like polymer after $10^{6}$ Monte Carlo iterations. (Left) $\beta^{-1}=0.1$ and (right) $
\beta^{-1}=1$. At low temperature the ribbon fluctuates around an underlying helical shape. The radius and the pitch of the helix are set by the persistence 
length $\ell_{p}$ and the wave-number $k$ and are proportional to $\beta a$ and $\beta^{1/2} a$ respectively.}
\end{figure}

This oscillatory behavior is reminiscent of spin chains with competing ferromagnetic and antiferromagnetic interactions. In these systems the oscillatory 
behavior of the two-point correlation functions denotes the existence of a {\em helical phase} in which the magnetization varies periodically in space. 
The critical point that divide the disorder phase and the helical phase is traditionally referred to as ``Lifshitz point'' \cite{RednerStanley:1977}. 
Analogously, an oscillating tangent-tangent correlation function in our polymer model implies the presence of an underlying helical structure that persists 
at any finite temperature even in absence of a spontaneous twist, i.e. so that the Lifshitz point here is $T=\infty$. In Fig. \ref{fig:ribbon} we show that our numerical
simulations capture this feature, due to the strong geometrical constraint of developability, i.e. the ribbon is  isometric to a flat strip at {\em any} temperature, and thus
bends when twisted. The notion of a Lifshitz point  has been used earlier  to described the low temperature phase of a ``railway-track'' polymer consisting of two parallel 
WLCs connected by a rigid bond \cite{Liverpool}, and is marked by an oscillatory binormal-binormal correlation function which results in a ``rod-kink'' structure in 
which the ribbon is at every point along its contour either twisted and unbent (rod) or bent and untwisted (kink). This is qualitatively different from our results where  it is
impossible for the ribbon to be twisted and unbent simultaneously, due to the fact that the constraint of developability is implemented exactly in Eq. \eqref{eq:sadowsky}, 
unlike in any previous attempts.

\begin{figure}[t]
\centering
\includegraphics[width=0.95\columnwidth]{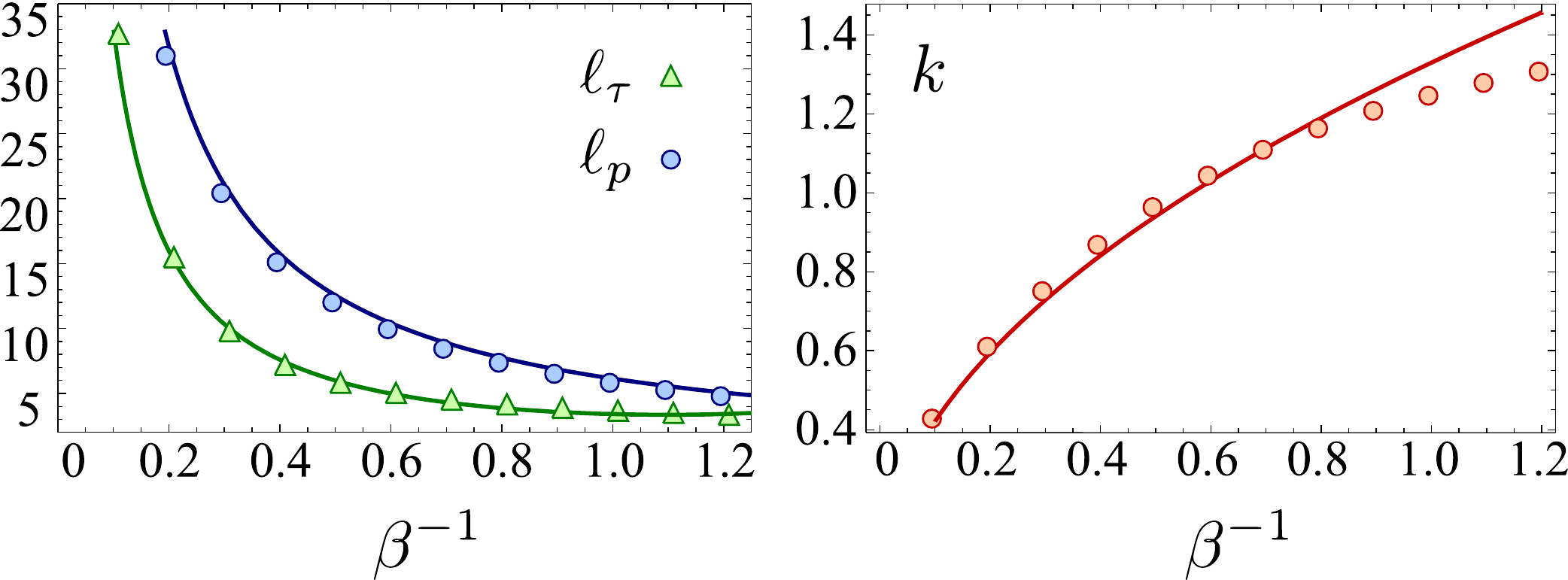}	
\caption{\label{fig:persistence-length}(Color online) Persistence length and torsional persistence length (left) as a function of $\beta^{-1}$ (in units of $a$). (Right) 
Tangent-tangent correlation function wave-number $k$ (in units of $a^{-1}$) as a function of $\beta^{-1}$. Dots corresponds to numerical data and lines to analytical predictions for the low temperature regime. For biopolymers of Young modulus $E\approx 100$ MPa, $h\approx 1$ nm and $w/a \approx 1$, the value of $\beta^{-1}$ in the plots spans the temperature range $T\in[0,\,10^{3}]$ K.}
\end{figure}

Next we give an analytical estimate of the three relevant quantities $\ell_{p}$, $\ell_{\tau}$ and $k$ in the low temperature regime. The partition function
for the ribbon is given by $Z=z_{1}^{N}$ with:
\begin{equation}\label{eq:partition-function1}
z_{1} = \int \frac{d\Omega}{4\pi}\,e^{-\frac{1}{2}\beta a^{2}\left(\kappa^{2}+2\tau^{2}+\frac{\tau^{4}}{\kappa^{2}}\right)}\,,
\end{equation}
with $d\Omega=d\phi\,d\theta \sin\theta$ the infinitesimal solid angle. The analytical calculation of thermal average is complicated by the coupling term 
between curvature and torsion, but this can be removed by rewriting the $\tau^{4}/\kappa^{2}$ term as a Gaussian integral of the form:
\begin{equation}
e^{-\beta a^{2} \frac{\tau^{4}}{2\kappa^{2}}} = \left(\frac{\beta\,a^{2}\kappa^{2}}{2\pi}\right)^{\frac{1}{2}}\int_{-\infty}^{\infty} dy\,e^{-\frac{1}{2}\beta a^{2} 
(\kappa^{2}y^{2}+2i\tau^{2}y)}\,.
\end{equation}
With this substitution, curvature and torsion are now decoupled and the single-node partition function reads:
\begin{multline}
z_{1} 
= \left(\frac{\beta}{\pi}\right)^{\frac{1}{2}}\int_{-\infty}^{\infty}dy\,\int \frac{d\Omega}{4\pi}\,(1-\cos\theta)^{\frac{1}{2}}\,\\e^{-\beta[\lambda(1-\cos\theta)+\mu(1-
\cos\phi)]}\,,
\end{multline}
with $\lambda=(1+y^{2})$ and $\mu=2(1+iy)$, so that the ribbon is equivalent to a wormlike polymer with finite but complex-valued bending and torsional stiffness. 
Calculating the solid-angle integral in the previous expression yields 
\begin{equation}
z_{1} = \left(\frac{\beta}{4\pi}\right)^{\frac{1}{2}}\\[7pt] 
\int_{-\infty}^{\infty} dy\, \frac{e^{-\beta \mu}I_{0}(\beta\mu)\,\gamma(\tfrac{3}{2},2\beta\lambda)}{(\beta \lambda)^{\frac{3}{2}}}\,,
\end{equation}
where $I_{0}$ and $\gamma$ are respectively the modified Bessel function of zeroth order and the incomplete gamma function \cite{AbramowitzStegun:1996}.  Analogously one can calculate the ensemble average $\langle\,\cdot\,\rangle = z_{1}^{-1}\int\frac{d\Omega}{4\pi}\,(\,\cdot\,)\,e^{-\frac{1}{2}\beta a^{2}(\kappa^{2}+2\tau^{2}+\tau^{4}/\kappa^{2})}$ which yields:
\begin{gather*}
\langle \cos\theta \cos m\phi \rangle = \left(\frac{\beta}{4\pi}\right)^{\frac{1}{2}}z_{1}^{-1} \int_{-\infty}^{\infty}dy\,e^{-\beta\mu}I_{m}(\beta\mu)\,f_{1}(\beta
\lambda)\,,\\[7pt]
\langle \sin\theta \cos m\phi \rangle = \left(\frac{\beta}{4\pi}\right)^{\frac{1}{2}}z_{1}^{-1} \int_{-\infty}^{\infty}dy\,e^{-\beta\mu}I_{m}(\beta\mu)\,f_{2}(\beta
\lambda)\,,
\end{gather*}
where $m$ is an integer in $\{0,1\}$ and
\begin{gather*}
f_{1}(\beta\lambda) = \frac{\gamma(\frac{3}{2},2\beta\lambda)}{(\beta\lambda)^{\frac{3}{2}}}-\frac{\gamma(\frac{5}{2},2\beta\lambda)}{(\beta\lambda)^{\frac{5}{2}}}\,,\\[7pt]
f_{2}(\beta\lambda) = \frac{3}{\sqrt{2}(\beta\lambda)^{2}}-\frac{\sqrt{\pi}\,(3+4\beta\lambda)}{4(\beta\lambda)^{\frac{5}{2}}}\,e^{-2\beta\lambda}{\rm erfi}(\sqrt{2\lambda\beta})\,,
\end{gather*}
with ${\rm erfi}$ the imaginary error function \cite{AbramowitzStegun:1996}. At low temperatures ($\beta\gg 1$), asymptotic expansions of the functions $I_{0}$, $\gamma$ and ${\rm erfi}$ yield
\begin{gather*}
\langle \cos\theta \cos m\phi \rangle \simeq 1-\frac{21(1-4m^{2})}{16\beta}+\frac{4(7-173m^{2})}{5+128\beta}\,,\\[7pt]
\langle \sin\theta \cos m\phi \rangle \simeq \frac{128\beta[40\beta-7(2+m^{2})]-99(1-4m^{2})}{12\sqrt{2\pi}\,\beta^{3/2}(5+128\beta)}\,,
\end{gather*}
from which we find the persistence lengths $\ell_{p}$ and $\ell_{\tau}$ 
\[
\ell_{p} = \left(\frac{75}{64}-\frac{25}{9\pi}\right)^{-1}\frac{Dw}{k_{B}T}, \qquad 
\ell_{\tau} = \frac{32}{5}\,\frac{Dw}{k_{B}T}\,,
\]
and the wavenumber $k$:
\[
k = \frac{5}{3}\sqrt{\frac{2}{\pi a}\frac{k_{B}T}{Dw}}\,.
\]
A plot of $\ell_{p}$, $\ell_{\tau}$ and $k$ versus temperature is shown in Fig. \ref{fig:persistence-length}. We see that as a consequence of the geometric coupling between curvature and torsion in a developable ribbon, the persistence length is a factor $\left(\frac{75}{64}-\frac{25}{9\pi}\right)^{-1}=3.476$ larger than that of a WLC at the same temperature. 
%

Our study has a set of clear experimental predictions for the equilibrium statistical mechanics of ribbon-like polymers where all three length scales are well separated. 
Firstly, the tangent-tangent correlation function exhibits an oscillatory decay at any finite temperature, implying an underlying helical structure even in absence of any natural  
twist at zero temperature. For a ribbon with finite extensibility, even if the helical phase disappears at high enough temperatures, the Lifshitz point should persist but is 
shifted to $T=T^{\ast}$, and the mean field estimate for the oscillatory wave number  $k \sim [k_B (T-T^{\ast})/Dwa]^{1/2}$
%
%
The helical structures that arise herein are reminiscent of those found in biopolymers and proteins, and suggests a natural generalization of our theory to include short range 
forces that will stabilize these structures. In the presence of forces, the geometrical structure of ribbons is expected to lead to qualitatively differences in the force-extension curves of these ribbons as well as in their motion, topics of future study. 
%

This work was partially supported by NSF through the Harvard MRSEC and the Brandeis MRSEC, by the Harvard Kavli Institute for Bionano Science \& Technology
and by the MacArthur Foundation.

\end{document}